# A Sunspot Catalog for the Period 1952-1986 from Observations Made at the Madrid Astronomical Observatory

A.J.P. Aparicio[1,3] • L. Lefèvre[2] • M.C. Gallego[1,3] • J.M. Vaquero[3,4] • F. Clette[2] • N. Bravo-Paredes[1] • P. Galaviz[4] • M.L. Bautista[1]

[1]Departamento de Física, Universidad de Extremadura, Badajoz, Spain

[2]Royal Observatory of Belgium, Brussels, Belgium

[3]Instituto Universitario de Investigación del Agua, Cambio Climático y Sostenibilidad (IACYS), Universidad de Extremadura, Badajoz, Spain

[4]Departamento de Física, Universidad de Extremadura, Mérida (Badajoz), Spain

**Abstract**

Sunspot catalogs are very useful for studying the solar activity of the recent past. In this context, a catalog covering more than three solar cycles made by the astronomers of the Madrid Astronomical Observatory in Spain (nowadays, the National Astronomical Observatory) from 1952 until 1986 has been recovered. Moreover, a machine-readable version of this catalog has been made available. We have recovered abundant metadata and studied the reliability of this dataset by comparing it with other sunspot catalogs.

**1. Introduction**

Traditional indices for long-term studies on solar activity, *i.e.*, sunspot number, group number and sunspot area, describe the general state of the Sun (Hoyt and Schatten, 1998; Balmaceda *et al.*, 2009; Clette *et al*., 2014; Hathaway, 2015; Carrasco *et al.*, 2016; Muñoz-Jaramillo and Vaquero, 2018). On the other hand, for studies that require a more detailed approach, sunspot catalogs are a more appropriate tool. They offer data about positions, areas and types of groups and spots, which makes it possible to track the evolution of these structures during their way across the solar disk (Aparicio *et al*., 2014; Carrasco *et al*., 2014). However, each catalog has its own time coverage, contents and format. In order to expand the time coverage, fill gaps, correct mistakes and have a homogeneous source of data, the availability and merging of multiple sunspot catalogs is needed (Lefèvre and Clette, 2014; Carrasco *et al*., 2015).

In recent years, an effort has been made to retrieve solar data from different observatories around the world including Coimbra (Carrasco *et al.*, 2018), Ebro (Curto *et al.*, 2016), Greenwich (Willis, Wild, and Warburton, 2016), Locarno (Cortesi *et al*., 2016), Kodaikanal (Mandal *et al*. 2017) and Valencia (Carrasco *et al.*, 2014). In the context of this collective effort, different datasets preserved in several libraries and archives related to the solar and magnetic observations made at the Madrid Observatory during more than one century are being digitalized and analyzed.

In particular, an observational program devoted to sunspots was carried out at the Madrid Astronomical Observatory from 1876 to 1986. A lot of metadata about this solar program can be found in López Arroyo (2004) and Aparicio *et al*. (2014). The first observations started in 1868, although the program was not implemented on a systematic basis until 1876. In 1986, the program ended and, as a result, there are sunspot counting and sunspot area records for the periods 1876-1896, 1906-1920, 1931-1933 and 1935-1986, which were digitized, analyzed and made available by Aparicio *et al*. (2014). In addition, there are sunspot catalogs for the periods 1914-1920 and 1952-1986. The former catalog (Aguilar's catalog) was digitized, analyzed and made available by Lefèvre *et al*. (2016), and the same task is performed for the latter catalog ("modern" catalog) in the present article.

Moreover, the astronomers of Madrid developed a program to calculate the solar constant using pyrheliometers (Antón, Vaquero, and Aparicio, 2014; Aparicio *et al.*, 2019), and, very recently, the geomagnetic measurements made in this observatory during the 19th century have been recovered and analyzed (Pro, Vaquero, and Merino-Pizarro, 2018). This article continues with this process of data recovery from one of the oldest scientific institutions in Spain. In the next sections, the "modern" sunspot catalog of Madrid Observatory (from 1952 to 1986) is recovered and analyzed.

## 2. Structure of the Catalogs

The "modern" catalog was published in an internal journal of the observatory called *Boletín Astronómico del Observatorio de Madrid* (BAOM). Issues were published quasi-annually containing mainly information about astronomical observations made in this observatory. One of the sections is devoted to solar activity. In there, depending on the year, there are different sub-sections with data about daily sunspot numbers and group counting (recovered by Aparicio *et al.*, 2014), the sunspot catalog described in this work, protuberances, filaments and chromospheric faculae, among others.

Each sub-section devoted to the modern sunspot catalog contains metadata (explanation about instruments, observers, methodology and data) and one year of the catalog organized in two tables. The first one contains daily data about positions and areas of sunspot groups (main catalog), and the second one contains the Waldmeier classification for each group (catalog with Zurich types). The main catalog is sorted by date and spans the period 1952-1986, whereas the catalog with Zurich types is sorted by numbering of the groups and only spans the period 1952-1956.

The solar observations were performed with a Grubb telescope (20 cm aperture and 3 m focal length) and a Herschel helioscope eyepiece. As a result, 25 cm diameter images were projected onto a screen. Then, the sunspots were drawn on templates. The data to elaborate the original catalog were obtained from those drawings. Figure 1 shows two pages of the original catalog (one of the main catalog and one of the catalog with Zurich types). The columns in the main catalog are:

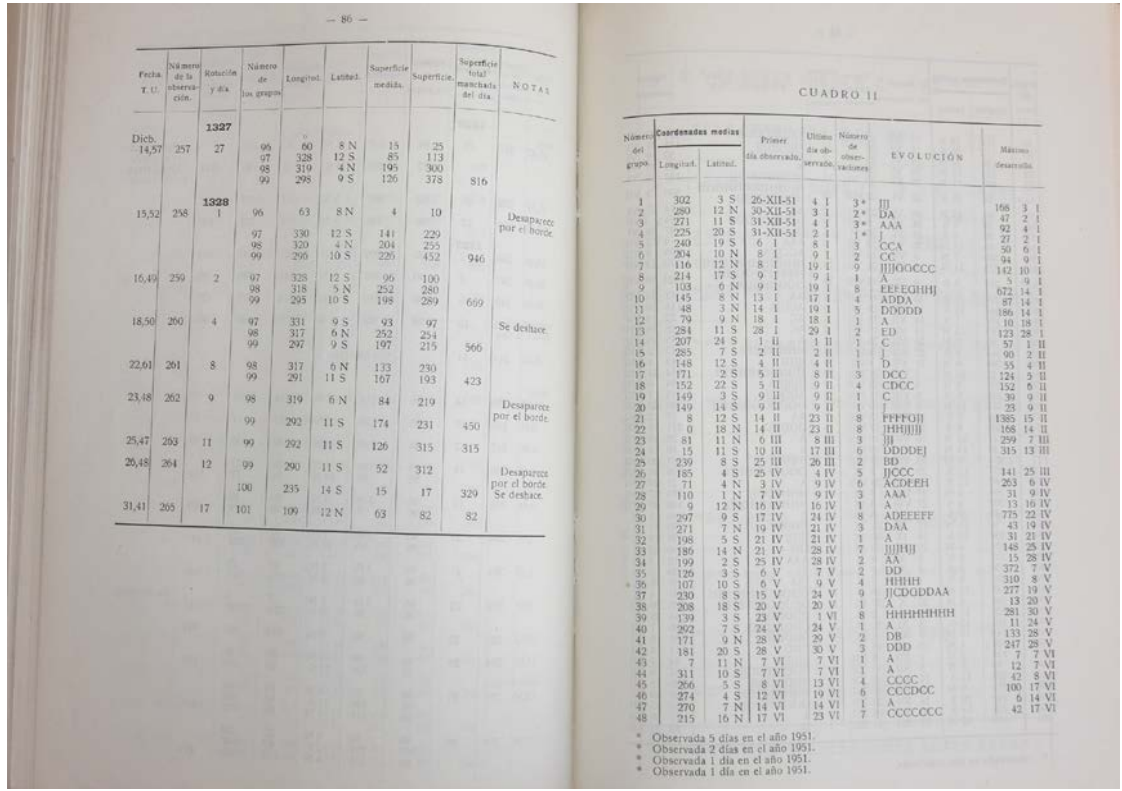

**Figure 1.** Example of two pages of the modern catalog (Boletín Astronómico del Observatorio de Madrid, 1953, IV (6), pages 86 and 87). Left: one page of the main catalog. Right: one page of the catalog with Zurich types.

i) Date: month and decimal day (in Universal Time).

ii) Number of the observation in the year.

iii) Solar rotation number and day.

iv) Number of the group. The first group observed is assigned number 1. The numbers are kept from one day to the next one, but the count starts from 1 every year.

v) Heliographic longitude in degrees.

vi) Heliographic latitude in degrees.

vii) Projected area in millionths of the solar disk (msd) divided by 2.

viii) Area corrected for foreshortening in millionths of the solar hemisphere (msh).

ix) Total sunspot area for the day (corrected for foreshortening).

x) Notes: the way the groups disappear.

The columns in the Zurich types catalog are:

i) Number of the group.

ii) Average coordinates. Left: Heliographic longitude in degrees. Right: Heliographic latitude in degrees.

iii) Date when the group was recorded for the first time.

iv) Date when the group was recorded for the last time.

v) Number of days with records of the group.

vi) Evolution of the group (Waldmeier classification).

vii) Sunspot area and date when the group reaches its maximum size.

In relation to the author of the catalog, some aspects have to be taken into account. On the one hand, there were a lot of observers who made the drawings to elaborate both sub-sections, the one with data about daily sunspot numbers and group counting (recovered by Aparicio *et al.*, 2014), and the other one with the modern sunspot catalog. On the other hand, the main observer of each year used to draw up the sunspot catalog. Table 1 lists the names of observers and authors of the catalog.

**Table 1.** Observers who made solar drawings and authors of the sunspot catalog of the Madrid Astronomical Observatory. The acronyms are: G = Gullón, M = Martín Lorón, L = Lagomacini, LA = López Arroyo, Li = Liñán, E = Encinas, R = Rodríguez, Lf = Lafuente, C = Claver, P = Pensado, A = Azcona, Ca = Castro, DP = De Pablo, E = Eliz, B = Balo, S = Sánchez, Gi = Gil, Ga = García, Cab = Cabañas, V = Vallejo. The number of drawings is in brackets.

| Year | Observers | Authors | Collaborators |
|---|---|---|---|
| 1952 | G (153), M (7), L (105) | G | |
| 1953 | G (147), LA (60), L (74) | G | LA, L |
| 1954 | G (56), LA (236), L (19) | G, LA | L |
| 1955 | G (40), LA (175), Li (5), L (20) | G, LA | Li, L |
| 1956 | G (56), LA (212), Li (8), L (5) | G, LA | Li, L |
| 1957 | G (5), LA (66), Li (1), L (26) | G, LA | Li, L |
| 1958 | G (2), LA (143), L (20) | G, LA | L |
| 1959 | G (11), LA (82), L (23) | G, LA | L |
| 1960 | LA (79), L (3) | G, LA | L |
| 1961 | G (9), LA (139), L (20) | G, LA | L |
| 1962 | G (28), LA (160), L (22) | G, LA | L |
| 1963 | G (53), LA (107), L (29) | G, LA | L |
| 1964 | G (60), LA (90), L (69) | G, LA | L |
| 1965 | G (227), LA (1), E (23), R (1) | G | |
| 1966 | G (215), LA (27), E (3), R (2), M (27) | G | |
| 1967 | G (279), Lf (1), R (6), C (46) | G | |
| 1968 | G (185), C (106), R (25) | G | Lf |
| 1969 | C (207), R (51), Lf (29) | P | Lf |
| 1970 | C (219), R (59), Lf (23) | P | Lf, A, C |
| 1971 | P (135), Ca (17), A (23), C (41), R (56) | P | |
| 1972 | P (56), A (111), R (62) | P | |
| 1973 | LA (61), DP (20), A (98), R (85) | LA | C |
| 1974 | LA (68), Ca (2), DP (11), R (165), C (14) | LA | C |
| 1975 | LA (66), DP (22), R (176), C (12), A (2) | LA | C |
| 1976 | LA (113), A (3), R (108), C (28) | LA | C |
| 1977 | LA (119), R (13), C (54), DP (20), E (11) | LA | C, E |
| 1978 | LA (116), E(45), C (52), B (19) | LA | C, E |
| 1979 | LA (92), E (55), S (12), C (28), B (7) | LA | C, E, Gi |
| 1980 | LA (89), Gi (16), E (34), S (7), C (64), Ga (6) | LA | C, Gi, E |
| 1981 | LA (103), C (34), Gi (67), E (2) | LA | Gi |
| 1982 | LA (67), Gi (89), S (6), B (18), E (1), Ga (1) | LA | Gi |
| 1983 | LA (56), Gi (110), S (6), A (1), Ga (2) | LA | Gi |
| 1984 | LA (15), V (7), Cab (19), Gil (128) | LA, Gi | |
| 1985 | Gi (147), S (3), Cab (11), V (8), Ga (1) | Gi, LA | |
| 1986 | Cab (2), Gi (78), V (1), S (39) | Gi | |

Several aspects of the catalog are worth noting. Firstly, the staff of the observatory recorded sunspot observations every day weather permitting. For this reason, there are days with no sunspots in the catalog. Secondly, the smallest reported area is 2 msd and 1 msh. Thirdly, the units of measurement for areas are not stated in the original catalog. It is stated that column 7 contains "sunspot area directly measured" and column 8 contains "sunspot area corrected for foreshortening". We used Equations 1 and 2 to infer the units of measurement by recomputing all the values of one column of area from the other as follows:

$$SA(msh) = \frac{SA(mm^2) 10^6}{2\pi R^2} \frac{1}{\cos(\arcsin\frac{r}{R})} \qquad (1)$$

$$SA(msh) = \frac{SA(msd)}{2} \frac{1}{\cos(\arcsin\frac{r}{R})} \qquad (2)$$

where SA(msh) is the sunspot area in msh, SA(mm$^2$) is the sunspot area in mm$^2$, SA(msd) is the sunspot area in msd, $r$ is the distance from the center of the group to the center of the disk, and $R$ is the radius of the image (12.5 cm). We found that Equation 1 does not work properly (even supposing that the units of measurement in column 7 could be larger or smaller than mm$^2$). However, Equation 2 does by taking column 7 as SA(msd)/2 (not as SA(msd)) and column 8 as SA(msh).

This can easily be confirmed by comparing the data with another available source for group areas. The Greenwich catalog (often called RGO catalog) contains areas in msd and in msh, and we can actually match day by day the different groups of both catalogs (see Figure 2). This comparison will be explained in depth in Section 4. Here we show only the relevant information. Figure 2 reveals that the relation between RGO individual group areas in msd and Madrid column 7 is double than the relation between RGO individual group areas in msh and Madrid column 8. Therefore, it proves that column 7 of the Madrid catalog corresponds to SA*(*msd*)*/2 (not SA(msd)) while column 8 of the Madrid catalog corresponds to SA*(*msh*)*.

Regarding the time coverage of the Madrid catalog, some global features are worth mentioning. The catalog presents 31782 lines and 6645 individual sunspot groups. Apparent time coverage is still spotty with a little below 39 % of days where there are no observations. Contrary to the data in Lefèvre *et al*. (2016), the days when there are no spots are differentiated from the days with no observations. Table 2 summarizes this information.

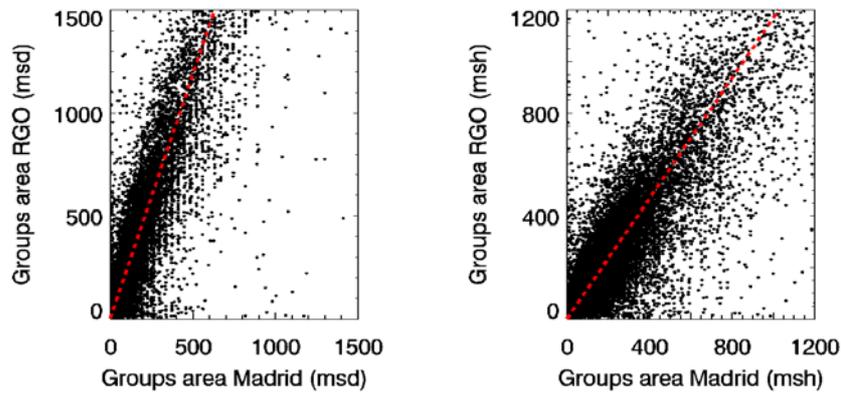

**Figure 2.** (Left) Relation between the areas of matched groups in millionths of the solar disk (msd) in the RGO catalog and column 7 of the Madrid catalog on the overlapping period. (Right) Same as (left) in millionths of the solar hemisphere (msh) in the RGO catalog and column 8 from the Madrid catalog. Dashed red lines show the combination of the results of six different fits: X, Y linear fit; Y, X linear fit; X, Y linear fit through 0; Y, X linear fit through 0; mean ratio of the areas; X, Y total least square fit, where X and Y represent the axes of the graphs. The combined fit equations (weighted mean and standard deviation) are $Y = (2.36 \pm 0.12) X$ for the left panel and $Y = (1.16 \pm 0.07) X$ for the right one.

**Table 2.** Summary of observations for years 1952-1986 in the Madrid Catalog

| | |
|---|---|
| ❖ First day | 02/01/1952 |
| ❖ Last day | 22/12/1986 |
| ❖ Theoretical number of days | 12774 |
| ➤ Days with no observations | 4963 |
| ➤ Days with observations | 7811 |
| ▪ Days with no groups | 1268 |
| ▪ Days with groups | 6543 |
| • Number of groups | 6645 |
| ❖ Time coverage | 7811/12774 (61 %) |

## 3. Errors in the Catalog

There are different ways to check for inconsistencies in the catalog. They depend on what kind of information is available. In this case, we have heliographic latitude and longitude as well as two types of areas for the individual groups (columns 7 and 8, see Section 2), and the total area of the groups observed on one day. This gives us mainly three possible checks:

i) Compute SA(msh) from SA(msd) and vice versa (by means of Equation 2) and compare to the actual values in the catalog.

ii) Compute the longitude from central meridian (LCM = $L - L_0$) from the heliographic longitude ($L$) and $L_0$ (input is the date). If LCM is not in the right range, then, most probably the longitude had an error.

iii) Compute the total daily area from the group areas.

Once we have computed these values, and detected a possible mismatch/problem, we have to check the origin. Some errors are the consequence of an error in another parameter/column. For example, if the longitude value is corrupted, the computation of the corrected area is affected. So i) or iii) might be the consequence of ii).

If the errors were due to a mistake during the digitization process, *i.e.* did not appear in the paper catalog, they were corrected straight away. If however, they appear in the same way in the original paper catalog, we can only use our best guess to correct the mistake. So we propose two versions of the catalog with flags: i) the original catalog with flags, and ii) the version where we propose corrections for the detected errors that are not due to faulty digitization. A description of the flags can be found in Table 3.

**Table 3.** Description of the different flags.

| FLAG # | Description | | Cause |
|---|---|---|---|
| 1 | LCM recalibrated, incompatible areas, area2 < area1 | $r > 0.9$ | Size/long |
| 2 | LCM recalibrated, incompatible areas, area2 < area1 | $r \leq 0.9$ | Size/long |
| 3 | Abs(LCM) > 90°, incompatible areas, area2 < area1 | | Longitude |
| 4 | Abs(LCM) ≤ 90°, incompatible areas, area2 < area1 | $r \leq 0.9$ | Size |
| 5 | Abs(LCM) ≤ 90°, incompatible areas, area2 > area1 | | Size |
| 6 | Total daily area ≠ sum of areas | | Size |

LCM: Longitude from central meridian; area1 and area2 are the columns 7 and 8 of the main catalog; $r$ is the distance from the disk center.

Of course, we can use comparison catalogs to detect errors or especially infer corrections for these errors. This is done is Section 4. The resulting correction propositions are a result of intra and inter comparison between available and overlapping sunspot catalogs.

## 4. Comparison with Other Sources

We can compare the Madrid catalog with information from other overlapping sunspot catalogs. The Debrecen catalog could be used, but considering the very limited overlap (DPD starts in 1974), and the fact that RGO should be limited to 1976, it has not been done here. We use the USET catalog (1940-now) and the RGO catalog (1876-1976/1986). Because of the differences in areas in RGO before and after 1976, we cut the RGO analysis in two parts around 1976.

### 4.1. Comparison with the USET Catalog

The USET catalog contains information on the groups of spots from March of 1940 to the present. The main features of this catalog are listed in Table 4.

Table 4. Summary of sunspot information in the USET catalog.

| Observations | March 1940- present |
|---|---|
| Quality of drawing | 1-5 (bad-perfect) |
| Dates | Year, month, day, hour, minutes, seconds |
| Positions | Heliographic latitude, longitude |
| Morphological type | McIntosh/Zurich |
| Number of spots | - |
| Latitude, Longitude | Both dipole ends |
| Area of group | Implementation in progress |

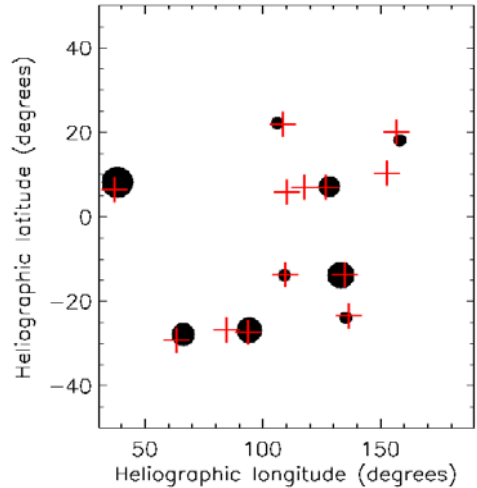

**Figure 3.** Synoptic map of the Sun on March 22, 1979. Black dots represent the data from the Madrid modern catalog while the red crosses represent the groups from the USET database.

The association process used in the present work to match groups from one catalog to another is described in Lefèvre *et al*. (2016). The groups in the Madrid catalog are compared to all the groups in the USET catalog. For a certain group in the Madrid catalog, the closest group (in the USET catalog) that is less than one day apart is taken, as long as the group meets a consistent distance criterion based on a preliminary distance analysis. Figure 3 shows a typical comparison of groups on the Sun's surface between the USET and Madrid catalogs. The distribution of the distance between matched groups is presented in Figure 4 with the same computation scheme that was applied in Lefèvre *et al*. (2016). From these two figures, it can be seen that the association process works properly. In addition, Figure 3 suggests that the USET catalog detects more groups than the Madrid one. In fact, if we compute for each day the difference between the number of groups reported in each catalog, it is interesting to note that there is this bias towards USET. 467 days show the Madrid catalog contains more groups than the USET one, 2300 days show the opposite, while 2048 days show an identical number of groups in both catalogs.

A.J.P. Aparicio *et al.*

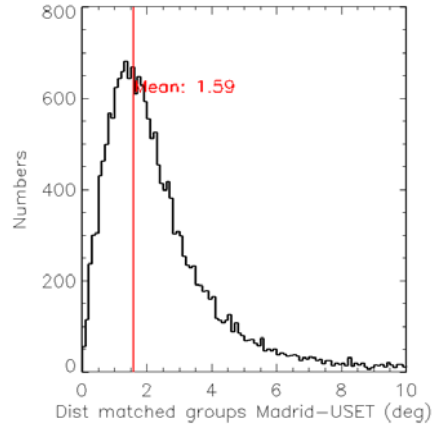

**Figure 4.** Distance between matched groups in Madrid and USET catalogs (bins of 0.1 degrees). The red line represents the mean value.

With respect to the groups areas, the computation is in progress at USET, within the context of a Belgian project, but is not advanced enough to be used at this time (it should be completed and quality analyzed by the end of 2021). However, we can use the Zurich types from the USET catalog and compare them to the types from the Madrid catalog.

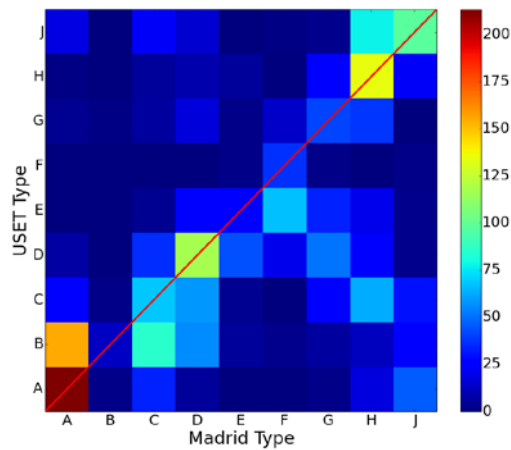

**Figure 5.** Two-dimensional histogram for Zurich types from Madrid (X axis) compared to Zurich type from USET (Y axis) for identified groups between the two catalogs. The red line represents the one-to-one correspondence, *i.e.* the USET and Madrid catalogs attributed the same type.

Figure 5 shows that we are mostly close to a one-to-one correspondence. However, the A, J or C, G and H types are less easy to classify here. For example, an H type group in the Madrid catalog corresponds to an H type group in the USET catalog, but also to a C-type group. As was mentioned in Lefèvre, Clette, and Baranyi (2011), the C-type classification suffers from a lack of determination making it sort of a "throw-in-everything" class, which could explain this behavior. Other examples of classification problems are A and J classes in both catalogs. As it is easy to confuse one with the other without evolution information, this is a classical misclassification and easily explains the mismatch between two catalogs. It is also interesting to note that the Madrid catalog contains very few B-types and that what would be called a B-type group in USET is often classified as A in Madrid. This shows the bias this classification can suffer from between different sources/observers.

4.2. Comparison with the Royal Greenwich Observatory Catalog (RGO)

A description of the RGO catalog can be found in Willis *et al.* (2013a,b). Data about sunspot groups from 1874 to 1982 are listed in this catalog. Nevertheless, after 1976 parameters are not consistent with the previous ones since they were obtained from the Boulder Solar Region Summary on a daily basis (Hathaway, 2015). As noted by this author, areas in RGO are 40 % smaller starting in 1977.

Figure 2 (right panel) shows the relation between group areas in the RGO catalog *versus* what we can find in the Madrid catalog for the 1952-1982 period. First of all, a high scatter is apparent. The figure was re-drawn taking |LCM| < 40° and |LCM| < 20° and the scatter remains. Therefore, groups near the limb do not cause the problem. The scatter is probably due to the use of different instruments, observers, observing conditions and especially different measurement methods. Secondly, if we analyze both periods around 1977 separately, we clearly see the dichotomy mentioned in the above paragraph. For the 5 years between 1977 and 1981, S(RGO) = (0.90 ± 0.06)·S(Madrid), while S(RGO) = (1.23 ± 0.07)·S(Madrid) before that. The ratio between these relations (taking error bars into account) is 1.35 ± 0.08, which is compatible with the 40 % decrease in RGO areas in 1977.

The association process used here have been the same as in the previous subsection. Figure 6 shows a typical comparison of groups on the Sun's surface between the RGO and Madrid catalogs, while Figure 7 shows the distribution of the distance between matched groups with the same computation scheme as in Figure 4.

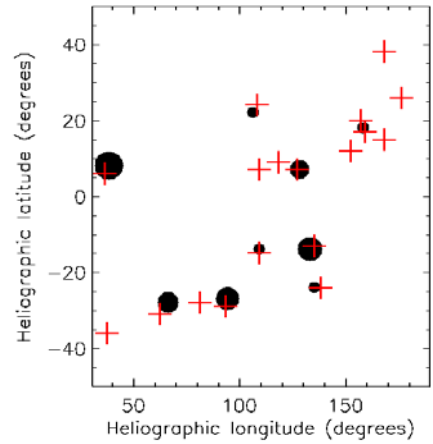

**Figure 6.** Synoptic map of the Sun on March 22, 1979. Black dots represent the data from the Madrid modern catalog while the red crosses represent the groups from the RGO database.

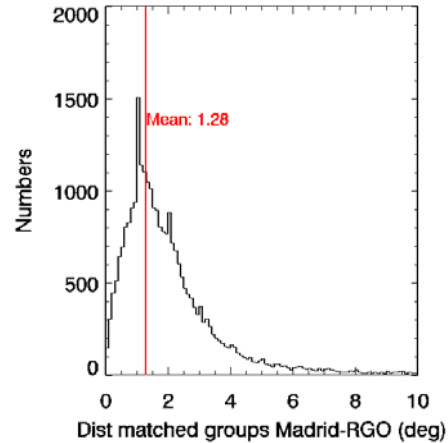

**Figure 7.** Distance between matched groups in Madrid and RGO catalogs (bins of 0.1 degrees). The red line represents the mean value.

A Sunspot Catalog from the Madrid Astronomical Observatory

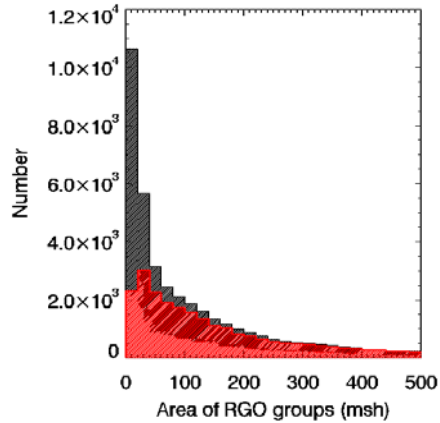

**Figure 8.** Size of RGO groups over the period 1952-1982 (bins of 20 msh). All the RGO groups are plotted in black. The RGO groups matched to a Madrid group are overplotted in red.

Figure 6 suggests that the Madrid catalog misses some groups. In order to check that, Figure 8 depicts the size of all the groups recorded in the RGO catalog *versus* the size of the RGO groups with an equivalent in the Madrid catalog. The histogram for the RGO groups with a counterpart misses a part of the low size distribution: *i.e.* below 40 msh, about 50 % of the groups have no equivalent, and below 20 msh the proportion increases to around 80 %. This proves that the Madrid catalog identifies a large part of the groups (larger than for the Aguilar catalog, Lefèvre *et al.*, 2016) but misses the lower part of the distribution. Although both (RGO and Madrid) catalogs register minimum areas of 1 msh, neither the RGO nor the Madrid catalog is complete below 5 msh (however, the RGO catalog is more complete that its Spanish counterpart).

Lastly, a comparison of the total sunspot area from the two observatories can be found in Aparicio *et al.* (2014). That work provides sunspot number, group sunspot number and sunspot areas of the Madrid Astronomical Observatory for the period 1876-1986. The series of sunspot areas for the period 1952-1986 was extracted from the catalog we present in this article. That series was compared to the Balmaceda composite (Balmaceda *et al.*, 2009), which, for the study period (1952-1986), mostly comes from RGO, and the rest is calibrated from it.

4.3. Group Number Comparisons

As it was mentioned in the previous subsection, Aparicio *et al.* (2014) provide a series of group sunspot number of the Madrid Astronomical Observatory for the period 1876-1986. Figures 9 and 10 present the group numbers extracted from the lists of the Madrid Astronomical Observatory compared to the number of groups extracted from the present catalog. They should be identical.

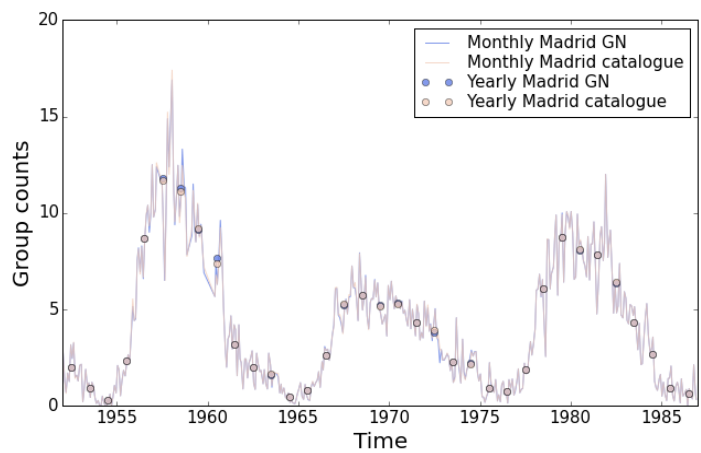

**Figure 9.** Monthly and yearly number of groups extracted from this catalog *versus* the group number (GN) from Aparicio *et al.* (2014). GN is divided by 12.08, since we are comparing raw number of groups.

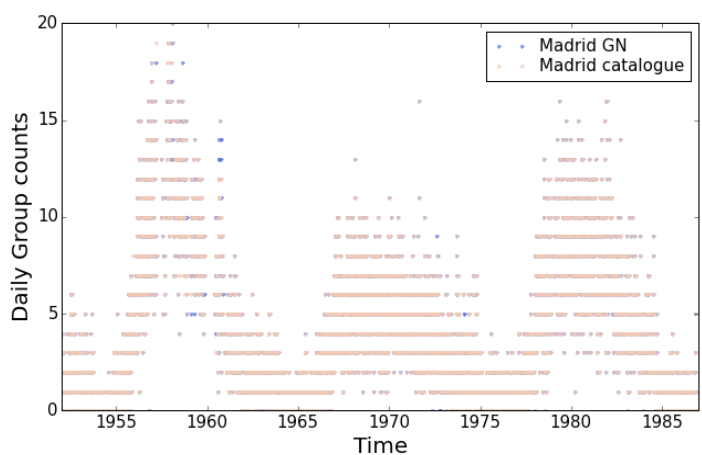

**Figure 10.** Daily number of groups from this catalog *versus* the daily group number from Aparicio *et al.* (2014) divided by 12.08.

Note in Figure 9 that in 1960, there seems to be a slight difference noticeable in the monthly and yearly values, but it is not significant.

## 5. Conclusions

In recent years, several sunspot catalogs have been recovered from archives and libraries of astronomical observatories with a long tradition observing the Sun. Lefevre and Clette (2014) highlighted the need to recover these catalogs and make inter-comparisons between them. The redundancy of data allows us to correct errors and to provide an accurate image of the behavior of sunspots during the last 150 years. In this work, an interesting catalog covering more than three solar cycles made by the personnel of the Madrid Observatory in Spain from 1952 until 1986 has been recovered. A machine-readable version of this catalog is available at HASO website.

The catalog of sunspots presented and analyzed here is of special relevance since it can serve as a link between the solar synoptic observation programs of the RGO from 1874 until 1976 and the NOAA/USAF Solar Observing Optical Network (SOON) since 1966. In fact, sunspot areas recorded by the RGO are about $40-50$ % larger than those measured by the SOON (see, for example, Foukal, 2014).

The comparisons made here between catalogs show the reliability of the data of the "modern" catalog of Madrid. No significant changes of non-solar origin have been detected in these data. In addition, we have abundant metadata, including the name of the observer for each record (a data often difficult to obtain in other catalogs made by several observers).

**Acknowledgements**

A.J.P. Aparicio thanks the Ministerio de Educación, Cultura y Deporte for the award of a FPU grant. L. Lefèvre's work is supported by the STCE and the BRAIN BR/165/A3/VAL-U-SUN grant. The authors have benefited from the participation in ISSI workshops. This work was partly funded by FEDER-Junta de Extremadura (Research Group Grant GR15137 and project IB16127) and from the Ministerio de Economía y Competitividad of the Spanish Government (AYA2014-57556-P and CGL2017-87917-P).

**Conflict of Interest**

The authors declare that they have no conflict of interest.

**Appendix**

Publications containing original data of the program of solar observations at the Madrid Astronomical Observatory are referenced in the following (the abbreviation BAOM stands for *Boletín Astronómico del Observatorio de Madrid*):

Gullón, E.: 1953, *BAOM* **IV**(6), 69–88.

Gullón, E.: 1954, *BAOM* **IV**(7), 83–99.

Gullón, E.: 1955, *BAOM* **IV**(8), 79–95.

Gullón, E., López Arroyo, M.: 1956, *BAOM* **V**(1), 77–109.

Gullón, E., López Arroyo, M.: 1957, *BAOM* **V**(2), 95–207.

Gullón, E., López Arroyo, M.: 1958, *BAOM* **V**(3), 25–66.

Gullón, E., López Arroyo, M.: 1959, *BAOM* **V**(4), 25–90.

Gullón, E., López Arroyo, M.: 1960, *BAOM* **V**(5), 17–56.

Gullón, E., López Arroyo, M.: 1961, *BAOM* **V**(6), 15–38.

Gullón, E., López Arroyo, M.: 1962, *BAOM* **VI**(1), 17–37.

Gullón, E., López Arroyo, M.: 1963, *BAOM* **VI**(2), 17–34.

Gullón, E., López Arroyo, M.: 1964, *BAOM* **VI**(3), 17–31.

Gullón, E., López Arroyo, M.: 1965, *BAOM* **VI**(4), 19–30.

Gullón, E., López Arroyo, M.: 1966, *BAOM* **VI**(5), 19–32.

Gullón, E.: 1967, *BAOM* **VI**(6), 25–53.

Gullón, E.: 1968, *BAOM* **VII**(2), 21–64.

Gullón, E.: 1969, *BAOM* **VII**(4), 21–68.

Pensado, J.: 1971, *BAOM* **VII**(6), 23–61.

Pensado, J.: 1972, *BAOM* **VIII**(1), 19–61.